\documentclass[12pt]{article}

\usepackage[english]{babel}
\usepackage[utf8]{inputenc} % allow utf-8 input
\usepackage{natbib}

\usepackage{amsmath}
\usepackage{amsthm}
\usepackage{amsfonts}
\usepackage{xcolor}
\usepackage{graphicx}

\newtheorem{theorem}{Theorem}[section]
\newtheorem{lemma}[theorem]{Lemma}
\newtheorem{assumption}[theorem]{Assumption}

\newcommand{\ind}{\perp\!\!\!\perp} 
\title{Assumption-free fidelity bounds for hardware noise characterization}
\author{Nicolo Colombo\footnote{{\tt nicolo.colombo@rhul.ac.uk}}}

\begin{document}

\maketitle
\begin{abstract}
In the Quantum Supremacy regime, quantum computers may overcome classical machines on several tasks if we can estimate, mitigate, or correct unavoidable hardware noise. 
Estimating the error requires classical simulations, which become unfeasible in the Quantum Supremacy regime.
We leverage Machine Learning data-driven approaches and Conformal Prediction, a Machine Learning uncertainty quantification tool known for its mild assumptions and finite-sample validity,  to find theoretically valid upper bounds of the fidelity between noiseless and noisy outputs of quantum devices. 
Under reasonable extrapolation assumptions, the proposed scheme applies to any Quantum Computing hardware, does not require modeling the device's noise sources, and can be used when classical simulations are unavailable, e.g. in the Quantum Supremacy regime. 
\end{abstract}
\section{Introduction}
\begin{figure}
    \centering
    \includegraphics[scale=.4]{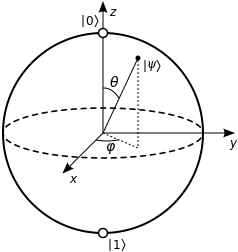}
    \caption{
    The Bloch sphere represents a qubit, $|\psi \rangle$ as the complex superposition of two states, $|\psi\rangle = \alpha |0\rangle  + \beta |1\rangle$, with $\alpha = \cos(\theta/2)$, $\beta = e^{i\phi}\sin(\theta/2)$, ${|0\rangle ={\bigl [}{\begin{smallmatrix}1\\0\end{smallmatrix}}{\bigr ]}}$, and ${|1\rangle ={\bigl [}{\begin{smallmatrix}1\\0\end{smallmatrix}}{\bigr ]}}$.
    Measuring a qubit can only produce two outcomes, $0$ with probability $\alpha^2$ and $1$ with probability $\beta^2$, and destroys the superposition (by setting either $\alpha$ or $\beta$ to 0). 
    If the qubit is entangled with another, measuring the first may affect the (unmeasured) state of the second.
    A quantum circuit consists of a series of matrix operators acting on $|\psi\langle$ before it is measured.
    }
    \label{figure bloch}
\end{figure}
The high potential of Quantum Computing (QC) has not manifested yet in real-world applications \citep{dasgupta2022assessing}.
Tasks where QC has proven scalability advantages \citep{arute2019quantum} require correcting errors induced by hardware noise.
Available Quantum Error Correction algorithms \citep{lidar2013quantum}, based on physics-informed Noise and Error Models (NEMs) that make strong assumptions on the underlying quantum process, are computationally expansive and become unreliable for deep circuits, i.e. circuits with many qubits. 
Existing Machine Learning (ML) methods to improve them \citep{harper2020efficient, canonici2024machine} often rely on unverifiable assumptions, e.g. the specific type of quantum noise affecting the device \citep{google2023suppressing, van2023probabilistic, acharya2024quantumerrorcorrectionsurface, mcewen2022resolving, thorbeck2023two}.
They also depend on the model architecture and training \citep{nguyen2021tensornetworkquantumvirtual, tindall2024efficient, patra2024efficient} and have theoretical success guarantees only in the limit of infinite samples \citep{zheng2023bayesian}.
Uncertainty estimation methods like \cite{caro2022generalization} or \cite{park2023quantum}mostly focus on Quantum ML applications and can be hardly extended to more general tasks.
In particular, it is unclear whether such Quantum ML approaches can directly estimate the output of a given $s$-qubit quantum device, which is a multivariate discrete distribution over ${\cal Y} = \{0, 1\}^s$.
Using black-box ML methods for data-driven characterizations of QC hardware noise is a relatively new but promising idea.
\cite{martina2022learning} and \cite{canonici2024machine} show that quantum noise is rich enough to provide a unique fingerprint of a QC machine.
The result is obtained from real hardware data but for small circuits (4 qubits).
Does the provided noise characterization still hold when the size of the circuits increases, e.g. when $s$ crosses the boundary of the so-called Quantum Supremacy regime \citep{preskill2012quantumcomputingentanglementfrontier}?
And can the characterization be used to provide statistically sound upper bounds for the quantum noise of classically intractable quantum hardware?

Conformal Prediction (CP) is a data-driven, scalable, and assumption-free uncertainty quantification technique.
CP's recent popularity relies on the increasing demand for trustworthy and reliable AI, with general applicability and mild assumptions.
Designed and commonly used to assess the uncertainty of ML systems in the finite-sample regime, CP is rarely exploited in AI-unrelated applications, e.g. to calibrate and characterize physical devices or complex natural processes.
In principle, CP's model-agnosticism may help quantify the output randomness of any partially observed physical system.
Compared to more standard statistical approaches, e.g. Bayesian inference, CP can be expected to be more robust because it does not require to guess or approximate the system's dynamics.
In QC, modeling the details of the data-generating process is particularly hard because of the unavoidable interactions with the environment and the complex-valued formalism of Quantum Mechanics, which is partially incompatible with the classical notion of probability \citep{feynman1951concept}.
Can a frequentist approach as CP overcome this challenge?

Using CP to evaluate the unmodeled parts of quantum hardware noise, we provide reliability guarantees for classically intractable QC systems.
Technically, we aim to test whether noise characterizations trained on small devices can be \emph{extrapolated} to devices and tasks that go beyond the Quantum Supremacy boundary.
Assuming hardware noise increases as a function of a device's features, e.g. its size and depth, we analyze the validity and efficiency of CP intervals when we calibrate and test the algorithm on circuits of different sizes.
A prior, usual CP validity is not guaranteed in this case because the output distributions of small and large devices may be non-exchangeable.

\subsection{Contribution}
\label{section contribution}

\paragraph{CP for physical systems.} Using CP beyond the evaluation of AI systems is new.
To our knowledge, there are no examples of CP-based schemes to directly estimate and quantify the uncertainty of physical systems.

\paragraph{Unpaired samples.} \cite{park2023quantum} use CP in the QC framework but focus on Quantum ML applications, which are trained to solve standard ML tasks, e.g. MNIST digit recognition.
Unlike \cite{park2023quantum}, the noisy and noise-free outputs considered here are associated with the \emph{predictions} and \emph{labels} of a standard ML task because they are unpaired.
Except for \cite{hu2024two}, the problem of evaluating unpaired samples, often referred to as the 2-sample problem, has not been addressed in the CP literature.
\cite{hu2024two}, likely the only existing work applying CP to a two-sample task, focuses on hypothesis testing for conditional distributions.
Our conceptual framework and ratio-based conformity score are similar but allow us to estimate continuous quantities (instead of a binary decision rule) and establish a clear link between the density-ratio conformity score and the target distribution divergences. 
Moreover, we focus on a specific case of conditional distributions where a heuristic ordering of the object space can be established in terms of a device's size and depth.

\paragraph{Non-exchangeability.}
In a QC real-world application, the outputs of classically tractable and intractable devices may be statistically different.
As we focus on small quantum machines when we train and calibrate the CP algorithms and intractable systems at test time, calibration and test sets are not exchangeable in our application.
Non-exchangeability issues have been widely addressed in the CP literature \citep{barber2023conformal}.
We propose two conceptually different new approaches inspired by the two common strategies used in CP to mitigate distribution shifts, sample reweighting \citep{bostrom2020mondrian, guan2023localized} and calibration training \citep{papadopoulos2008normalized, colombo2020training, colombo2024normalizing}. 
The proposed approaches are general and apply to any situations where training, calibration, and test sets come from different realizations of a smooth multivariate meta-distribution, e.g. when $P(Y|X = x) \neq P(Y|X = x')$.
\paragraph{Sample reweighting.}
In the QC setup,  non-exchangeability can be expected to vary as a function of specific feature summaries, e.g. the classical computational complexity required to simulate a system.
Our first mitigation approach consists of selecting training and calibration samples after ranking the data according to their complexity.
The strategy can be viewed as an extreme application of the Mondrian CP algorithm of \cite{bostrom2020mondrian}.
In particular, it is different from the distribution drift setup, e.g. \cite{gibbs2021adaptive}, where predictions become increasingly accurate as time passes (allowing asymptotic regret bounds).
We show how choosing the ranking function and a threshold reduces the expected undercoverage of the prediction intervals.

\paragraph{Calibration training.}
\cite{papadopoulos2008normalized} proposed to rescale the conformity scores with a pre-trained function of the inputs.
While CP-aware training of the prediction model has become common in the literature \cite{colombo2020training, bellotti2020constructing, stutz2021learning}, trainable conformity scores have been less explored.
In our scheme, we train a shift function that automatically compensates for the distribution drift.
Under idealized assumptions, we show that training the shift model reduces the total variation between calibration and test distributions, which is proportional to the validity gap defined in \cite{barber2023conformal}.

\paragraph{A general scheme.}
The proposed procedure applies beyond the QC framework and can quantify the accuracy of image and text generation models, where the output is also high-dimensional and possibly discrete.
Existing CP schemes in the image-generation framework are restricted to image-reconstruction tasks, where CP can be applied pixel-wise by comparing the model output and the original pixels. 
Bounding the overall distribution distance overcomes this limitation.
Moreover, our approach does not require designing expansive multiple-output CP approaches like \cite{messoudi2021copula}.

\subsection{Related Work}

\paragraph{CP for generative models}
Our work is related to the problem of applying CP to multivariate generative models for two reasons:
i) The output space is high-dimensional. 
ii) The predictor is an analytically unavailable \emph{generative distribution} from which we can only sample.
\cite{campos2024conformal} review recent CP approaches for Large Language Models. 
\cite{kutiel2023conformal} focus on visualization and use image masks to represent the uncertainty of an image regressor model.
\cite{belhasin2023principal} attempt to capture spatial correlations when computing CP per-pixel intervals but restrict to image restoration applications. 
We are unaware of approaches combining CP and the divergence measures used to train generative models, e.g. the Frechet Inception Distance (FID) or Kernelized Wasserstein divergence.
A combination of density ratio estimation and CP has appeared in \cite{tibshirani2019conformal} and \cite{hu2024two}.

\paragraph{Paired and unpaired samples}
Our approach is related to CP under random effects \cite{dunn2018distribution}.
Similar to here, CP intervals are obtained from the samples of an underlying black-box generator but are not explicitly related to distribution distances. 
\cite{ghosal2023multivariate} provide a conformal algorithm for when the inputs and outputs are distributions.
In their setup, however, each sample corresponds to a label,  i.e. attributes and labels are paired.
Outside the CP framework, techniques to address the 2-sample problem have been proposed in the statistical literature since 1939 \citep{smirnov1939estimation} and mostly in the non-parametric framework \citep{gretton2012kernel}.
The only connection between CP and 2-sample tasks is a recent work on hypothesis testing \citep{hu2024two} (see Section \ref{section contribution} for a technical comparison).

\paragraph{CP for QC}
In QC, CP has been used only in \cite{park2023quantum}, which is a direct application of Probabilistic CP \cite{wang2022probabilistic}.
In \cite{park2023quantum}, the multiple-output setup is avoided because high-dimensional observables can be handled by performing several separate measurements \cite{caro2022generalization}. 
The argument in \cite{caro2022generalization} relates to the generalization properties of Quantum ML algorithms.
It is unclear whether this is a more fundamental property of Quantum Computing devices.

\paragraph{Quantum Error Mitigation}
The problem of hardware noise is handled with QEM \citep{cai2023quantum}. 
Machine Learning (ML) methods have been used for QEM since 2020. Unlike us, ML-QEM often focuses on inferring distribution summaries, depends on the underlying error models and requires changing the target circuit \citep{harper2020efficient, liu2020reliability, strikis2021learning, wang2022probabilistic,canonici2024machine, adeniyi2025adaptive}.

\section{Methods}

\subsection{The Bhattacharyya Coefficient (BC)}
The output of small QC hardware can be reproduced using classical computers, i.e. a \emph{simulator} where all quantum operations are noise-free. 
In this case, we can compare the noisy outcomes of the hardware with the corresponding ideal output.
Let the ideal and noisy outputs be $Y_m \in \{0, 1 \}^s$ and $\hat Y_m \in \{0, 1 \}^s$, where $s$ is the number of measured qubits and $m = 1, \dots, M_{shots}$ indexes a single execution (shot) of the quantum machine.\footnote{$Y_1, \dots, Y_M$ and $\hat Y_1, \dots, \hat Y_M$ should be treated as random variables because of the intrinsic Quantum Mechanics non-determinism.}
$Y_m$ and $\hat Y_m$ can only take binary real values because a qubit collapses into a $\{ 0, 1\}$ when it is measured.
When $s$ qubits of a circuit are measured the noisy and ideal outputs are a series of binary strings, interpreted as i.i.d. draws from two underlying multivariate distributions, $P_Y$ and $P_{\hat Y}$, with support $\{0, 1\}^s$. 
As the noisy device and the classical simulator run independently, there is no relationship between single samples from $P_Y$ and $P_{\hat Y}$.
For simplicity, we use the same index for the samples from $\hat P_{Y}$ and $P_{\hat Y}$, even if a realization of $Y_m$ is \emph{not} the \emph{label} of a realization of $\hat Y_m$.
In this setup, hardware noise is proportional to any distribution distance between $P_{Y}$ and $P_{\hat Y}$.

The \emph{fidelity} of two quantum states is a quantification of their similarity, defined as the probability of identifying one state as the other.
Its classical counterpart, for discrete random variables, is the Bhattacharyya Coefficient  \citep{Bhattacharyya1946measure}, 
\begin{align}
\label{bhattacharyya coefficient}
{\rm BC}(Y, \hat Y) = \sum_{y \in {\cal Y}} \sqrt{P_Y(y) P_{\hat Y}(y)}
\end{align}
The BC is related to the Hellinger distance, 
$$d_{\rm H}^2(Y, \hat Y) =  \frac12\sum_{y\in {\cal Y}}\left (\sqrt{P_Y(y)} - \sqrt{P_{\hat Y}(y)}\right)^2$$ 
and the Total Variation distance, 
\begin{align}
    \label{total variation}
    d_{\rm TV}(Y, \hat Y) = \sum_{y \in {\cal Y}} \| P_Y(y) - P_{\hat Y}(y) \|
\end{align}
In particular, $1 - {\rm BC} = d_{\rm H}^2 \leq d_{TV} \leq \sqrt 2 \ d_{\rm H}$.
Inequalities involving the Kullback–Leibler divergence, $d_{\rm KL} = {\rm E} P_{\hat Y} \log\frac{P_{Y}}{P_{\hat Y}}$, can be found in \citep{sason2015reverse}.

\subsection{Large-scale devices}
A qubit can assume a dense infinity of states (all points in the Bloch sphere represented in Figure \ref{figure bloch}) and is described by a complex-valued wave function, $\psi$, usually encoded as a vector in an infinite-dimensional Hilbert space.
Simulating a quantum circuit requires computing the transformations induced by the circuit's logical gates on $\psi$. 
Classical simulations become infeasible when the number of interacting qubits in the circuit increases.
The \emph{Quantum Supremacy regime} is defined by the set of tasks that can only be completed in finite time by quantum devices, i.e. would have nearly infinite complexity on a classical machine \citep{preskill2012quantumcomputingentanglementfrontier}.
Current quantum computers are far beyond this limit \citep{decross2024computationalpowerrandomquantum}.
Intuitively, however, the boundary has no physical meaning and depends only on the available (classical) computational power compared with the circuit's complexity, depth, and size, i.e. the number of interacting qubits. 
This argument justifies  
\begin{assumption}
\label{assumption quantum supremacy}
    A quantum system's behavior may depend on its physical features, the environment in which is run, and its initial state, but not on whether it can be simulated on a classical machine.
\end{assumption} 
The requirement informally guarantees we can extrapolate a device noise characterization across the boundaries Quantum Supremacy regime.
Under Assumption \ref{assumption quantum supremacy} we can model the noisy output of a QC device as
\begin{align}
\label{noisy outputs}
    \hat Y \sim Y + \varepsilon_{noise}(Z), \quad Z = \phi(\hat Y) 
\end{align}
where $Z \in {\cal Z}$ is a general set of features describing the device and can be extracted from $P_{\hat Y}$ through a suitable feature map, $\phi$.
If $\varepsilon_{noise}$ is smooth enough \cite{buritica2024progression}, we can use ML to \emph{extrapolate} the gap between the ideal and noisy outputs of a classically intractable device by comparing the ideal and noisy outputs of smaller devices.
In practice, knowing or modeling the physical data-generating process underlying \eqref{noisy outputs} is challenging, especially because $\varepsilon_{noise}$ depends on partly unpredictable interactions between a quantum machine and the environment or the hardware defects.
In this sense, black-box ML systems like CP are ideal because they rely on minimal assumptions and apply to any black-box input-output system.
Assuming $\varepsilon_{noise}(Z) = \varepsilon_0 g(Z)$, where $\epsilon_0$ depends on the quantum machines where the circuit is run but not on the circuit size, we show how to adapt a standard Split-CP algorithm to produce prediction intervals that are valid even when calibration and test set do not come from the same distribution.

\subsection{Empirical BC approximation}
\label{section estimating}

\begin{figure}
    \centering
    \includegraphics[scale=0.3]{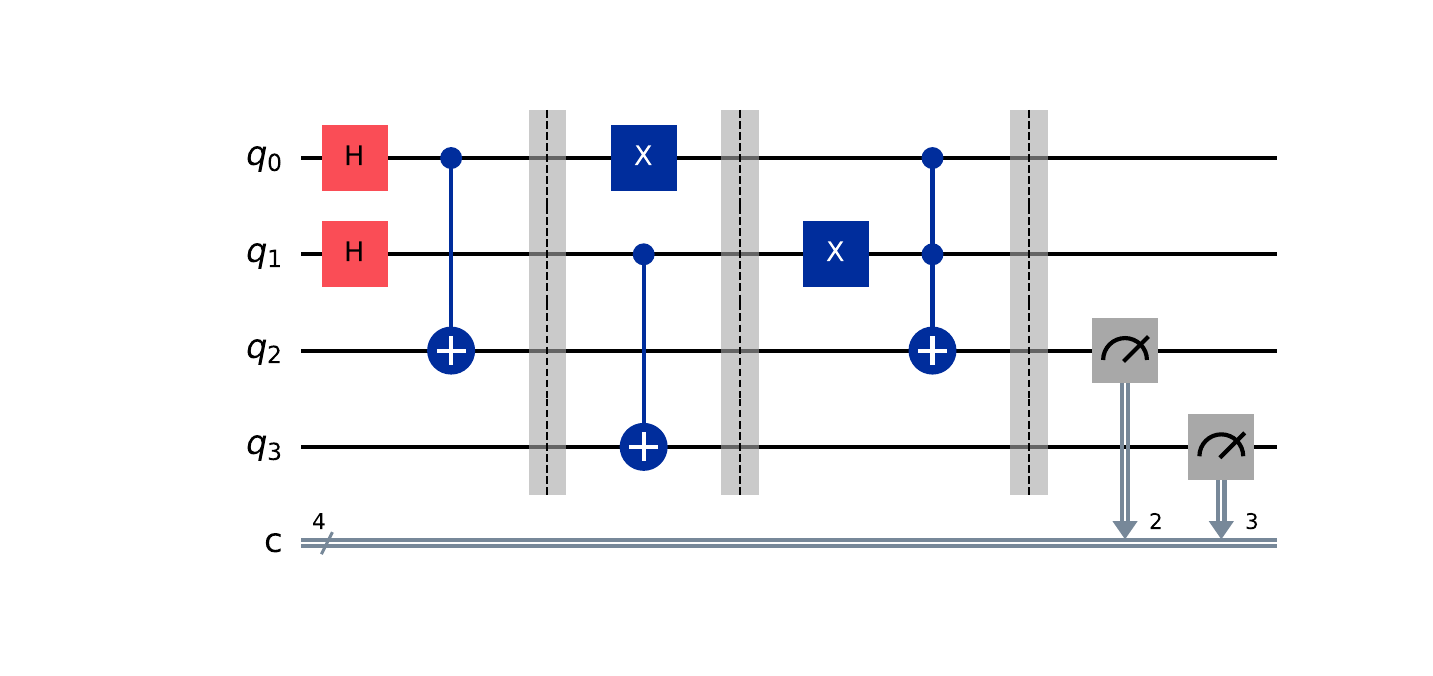}
    \caption{The modular 4-qubit quantum circuit of  \cite{martina2022learning}. 
    The 3-gate structure is repeated 3 times to increase the circuit depth.
    The top horizontal lines and blocks represent the device's interacting qubits and logical gates.
    The bottom line represents the device output.
    In this case, only 2 qubits are measured, $q2$ and $q3$, making each run produce a 2-digit binary string, e.g. $Y_{m}, \hat Y_{m} \in \{ {\tt 00}, {\tt 01},{\tt 10}, {\tt 11} \}$.
    The first two blue 2-qubit links represent CNOT gates and the third 3-qubit is a Toffoli gate. 
    See \cite{barenco1995elementary} for a formal definition. 
    The red boxes are Hadamard gates, used to prepare the unobserved qubits into an equal superposition state $\frac12(|0\rangle + |1\rangle)$.
    By convention, the initial state of all qubits is $|0\rangle$.
    The grey vertical lines represent 4-qubit synchronization. 
    }
    \label{figure walker}
\end{figure}

The dimensionality of the $n$-th distribution's support, $s_{n}$, is the number of qubits measured in the $n$-th circuit. 
In Figure \ref{figure walker}, $s_n = 2$ because only $q_2$ and $q_3$ are measured.
As the number of qubit measurements grows, e.g. for $d>5$, computing the BC explicitly becomes expansive or unfeasible because of the sum over all possible output configurations, $\sum_{y \in {\cal Y}_n} \sim O(2^{s_{n}})$.
Moreover, the explicit form of the distribution densities, $P_{\hat Y}$ and $P_{Y}$, is usually unknown and depends on the quantum dynamics of the qubits inside the circuit with and without hardware noise.
While Quantum Mechanics and the circuit layout may be used to obtain a theoretical estimate of $P_{Y}$, modeling $P_{\hat Y}$ would require including system-specific Noise Error Models for all device disturbances.
A data-driven option is to estimate the densities from the available samples using flexible enough ML algorithms, e.g. deep models or Kernel Density Estimation (KDE).
Nonparametric estimators like KDE are assumption-free but rely on the careful choice of the bandwidth and become unstable in high dimensions. 
Parametrized models, e.g. Neural Networks, may be misspecified and never capture substantial details of the distributions (even for an infinite sample size).
Our estimation strategy is hyperparameter-free and independent from the dimensionality of the distribution support. 
The idea is to replace straightforward density estimations with a more constrained but stable density-ratio estimation \cite{sugiyama2012density}.  
The need to estimate a density ratio is shared with other works, e.g. \cite{tibshirani2019conformal} or \cite{hu2024two}, and is closely related to CP conditional validity (see Section \ref{section extrapolation} for more details).
Our motivation for using a density-ratio estimator comes from a direct manipulation of the conformity score, $A= {\rm BC}(Y, \hat Y)$.
More explicitly, 
\begin{align}
        {\rm BC} &= \sum_{y \in {\cal Y} } \sqrt{P_{Y}(y) P_{\hat Y}(y)} = \sum_{y \in {\cal Y} } P_{\hat Y}(y) \sqrt{\frac{P_{Y}(y)}{P_{\hat Y}(y)}}\\
        &\sim \frac{1}{M_{shots}}\sum_{m=1}^{M_{shots}} \sqrt{\frac{p_{Y}(\hat Y_{m})}{p_{\hat Y}(\hat Y_m)}}  = {\rm E}_{\hat Y} \sqrt{\hat r(\hat Y)}
\end{align}
where $\hat r(\hat Y) \sim \frac{p_{Y}(\hat Y_{m})}{p_{\hat Y}(\hat Y_m)}$ is an empirical estimation of the ratio between the two densities evaluated and ${\rm E}_{\hat Y}(\hat r(\hat Y))$ its empirical expectation obtained from the available noisy samples, $\{Y_{n1}, \dots, Y_{nM_{shots}}\}$.
We refer to \cite{sugiyama2012density} for a review of the theoretical and empirical advantages of estimating $\hat r$ instead of the two densities separately.

\subsection{Conformal upper bounds}
Assume we have a set of $N$ classically tractable circuits and run them on a given QC machine.
Let $n=1, \dots, N$ index these $N$ circuits and $m=1, \dots M_{shots}$ the i.i.d. measurements from each circuit, i.e. the execution outputs, and $Y_{nm}$ and $\hat Y_{nm}$ the $m$-th noiseless and noisy measurements obtained from the $n$-th circuit.
As the dimensionality of a circuit's output depends on the number of measurements, we do not require $Y_{nm}$ and $Y_{n'm}$, to have the same dimensionality.
For example, we may have $Y_{nm} \in \{0, 1 \}^{s_{n}}$ and $Y_{n'm} \in \{0, 1 \}^{s_{n'}}$ with $d_n \neq d_{n'}$.
Naturally, we require different runs from the same circuit to have the same dimensionality, i.e. $Y_{nm} \in \{0, 1\}^{s_n}$ for all $m=1, \dots M_{shots}$).

Let $N+1$ index a test circuit that we can only run on the noisy device and $\hat P_{\hat Y_{N+1}}$ the corresponding output distribution. 
The prediction interval on the BC between $\hat Y_{N+1}$ and $Y_{N+1}$ produced by a Split-CP algorithm is an upper bound obeying 
\begin{align}
\label{marginal validity}
        &{\rm Prob}\left( {\rm BC}(Y_{N+1}, \hat Y_{N+1}) \geq Q_{\alpha}\right) \geq 1 - \alpha \\
        &Q_\alpha = \sup_q \left\{ \sum_{n=1}^N {\bf 1}(BC(Y_n, \hat Y_n) \geq q) \geq n_\alpha\right\}  \\
        & n_\alpha = \lceil(1 - \alpha)(N + 1)  \rceil 
\end{align}
The bound \emph{marginal validity} holds when the data generating distribution of the calibration and test sets, $\{(Y_n, \hat Y_n)\}_{n=1}^N$ and $(Y_{N+1}, \hat Y_{N+1})$, are exchangeable.  
In this case, \eqref{marginal validity} is guaranteed by the definition of the empirical quantile of exchangeable samples. 
See Lemma 1 of \cite{tibshirani2019conformal} for a simple proof of the quantile lemma and \cite{vovk2005algorithmic} or \cite{angelopoulos2021gentle} for a smooth introduction to CP.

Central to any CP algorithm is a scoring function, or \emph{conformity score}, $A$, which measures the quality of an output given the corresponding label.
In our case, 
\begin{align}
    A_n = {\rm BC}(\hat Y_{n}, Y_n), \quad n=1, \dots N
\end{align}
In practice, \eqref{marginal validity} is obtained by finding an upper bound for $A_{N+1}(Y) = {\rm BC}(\hat Y_{N+1}, Y)$, where $Y$ represent any $s_{N+1}$- dimensional random variable within distance $Q_\alpha$ from the unknown ideal output, $Y_{N+1}$.

\subsection{BC conformal extrapolation}
\label{section extrapolation}
Let $Z_1, \dots, Z_N$ and $Z_{N+1}$ be the features of $N$ quantum circuits that can be simulated on a classical machine and a large circuit that can not.
The $N$ tractable circuits are run on quantum hardware and a classical simulator to produce $\hat Y_{nm}$ and $Y_{nm}$, where $n=1, \dots, N$ and $m=1\dots, M_{shots}$.
The ${N+1}$-th circuit can not be simulated classically but can be run on the quantum hardware to obtain $\hat Y_{(N+1) \ m}$, $m=1\dots, M_{shots}$.
As the $(N+1)$-th circuit may be much more noisy than the others the conformity scores $A_{n}={\rm BC}(Y_{n}, \hat Y_{n})$, $n=1, \dots, N$ and $A_{N+1}={\rm BC}(Y_{N+1}, \hat Y_{N+1})$ may be non-exchangeable.
In this case, \eqref{marginal validity} does not hold and should be replaced by \citep{barber2023conformal}
\begin{align}
\label{approximate validity}
        &{\rm Prob}\left( {\rm BC}(Y_{N+1}, \hat Y_{N+1}) \geq Q_{\alpha}\right) \geq 1 - \alpha - {\rm gap}\\
        & {\rm gap} = \frac{1}{N+1}\sum_{n=1}^{N} d_{\rm TV}(A_{n}, A_{N+1}) \\
        &Q_\alpha = \sup_q \left\{ \sum_{n=1}^N {\bf 1}(A_n \geq q) \geq \lceil(1 - \alpha)(N + 1)  \rceil \right\} \\
        & A_n = {\rm BC}(Y_{n}, \hat Y_{n})
\end{align}
The practical use of the bound is limited because we cannot estimate $d_{\rm TV}(A_{n}, A_{N+1})$, $n=1, \dots, N$, without knowing all noise sources of the test device or having samples from the noiseless outputs, $Y_{N+1}$.
Two orthogonal general strategies have been proposed to mitigate similar non-exchangeability issues, sample reweighting, e.g. the Mondrian Conformal Prediction algorithm of \cite{bostrom2020mondrian} and the localization approaches of \cite{tibshirani2019conformal} and \cite{guan2023localized}, and calibration training, where the calibration score function is trained to account for possible object-conditional variability. 
\footnote{The latter approach was initiated by \cite{papadopoulos2008normalized} in 2028 but has been rarely exploited fully or investigated further. See \cite{colombo2024normalizing} for a more recent example.}.
These and follow-up works focus on establishing CP-conditional validity instead of addressing general shifts between the calibration and test distributions.
Replacing CP marginal validity \eqref{marginal validity} with an exact or approximate input-conditional version of it is challenging and, in general, practically unachievable \citep{vovk2012conditional}.
Assuming there exists a feature map, $Z = \phi(\hat Y) \in {\cal Z}$, defined for any $\hat Y$, i.e. for any QC device, estimating an upper bound of the BC between the noisy and noiseless outputs of classically intractable circuits would be equivalent to finding a quantile function $Q:{\cal Z}\otimes (0, 1) \to [0, 1]$, such that 
\begin{align}
\label{conditional validity}
    &{\rm Prob}\left( {\rm BC}(Y_{N+1}, \hat Y_{N+1}) \geq Q(z, \alpha) | Z_{N+1} = z \right)  \geq 1 - \alpha 
\end{align}
where $Z_n = \phi(\hat Y_n)$ are the features of the $n$-th device.
In particular, \eqref{conditional validity} would automatically provide valid noise quantification for large devices by setting $z = Z_{N+1}$ at test time.
In the Quantum Supremacy regime, however, this is impossible because circuits with $z = Z_{N+1}$ are classically intractable and cannot be used to calibrate the intervals.
More generally, estimating the empirical quantile of the conditional distribution for any $z \in {\cal Z}$ when ${\cal Z} \subseteq {\mathbb R}^d$ is impossible if $N < \infty$ because the probability of having $N$ samples with $z = Z_{N+1}$ in the calibration set is zero. \citep{vovk2012conditional}.
Here we focus on our specific setup and approximate conditional validity based on two observations
\begin{enumerate}
    \item We may be able to extract from $\hat Y$ an \emph{ordinal feature}, $S = \phi(\hat Y)$, such that
    \begin{align}
    d_{\rm TV}(BC_n, BC_{n'}) \propto f(|S_n - S_{n'}|), \quad S_n=\phi(\hat Y_n), \quad \partial_t f(t) > 0
\end{align}
For example, $S$ may be a weighted sum of a circuit's depth and number of interacting qubits. 
\item 
Since the object attributes are distributions, we can infer an arbitrarily high-dimension feature map, $\phi(\hat Y) \in {\mathbb R}^q$, such that $BC(Y_n, \hat Y_n) \approx w^T \phi(\hat Y)$ obeys 
\begin{align}
    \tilde A(\hat Y) = BC(Y, \hat Y) - w^T \phi(\hat Y) \sim P_{\tilde A} \quad  \text{for all } \hat Y
\end{align}
for some $w \in {\mathbb R}^q$\footnote{More formally, $\tilde A$ is such that the joint distribution of $\tilde A$ and $\hat Y$ factorizes, that is $P_{\tilde A \hat Y}=P_{\tilde A} P_{\tilde \hat Y}$  }  
\end{enumerate}
We formalize these assumptions to derive two non-exchangeability mitigation algorithms in Sections \ref{section calibration selection} and \ref{section calibration training}. 
In the experiments, we combine the algorithms and test them empirically against a baseline CP algorithm, i.e. the standard Split CP approach without non-exchangeability mitigation.

\subsubsection{Sample selection with an ordinal feature}
\label{section calibration selection}

The idea is to rank the calibration samples according to an ordinal feature extracted from the output distributions.
Assuming the test samples rank higher, we show that discarding low-rank calibration samples may reduce the validity gap in \eqref{approximate validity}.
In this section, we make the following
\begin{assumption}
\label{assumption ordinal}
Let $Y, \hat Y$ and $Y', \hat Y'$ be any two noiseless and noisy distributions of two devices, $A = BC(Y, \hat Y)$, and $A' = BC(Y', \hat Y)$.
There exists a one-dimensional score, $\phi(\hat Y) \in {\mathbb R}_+$, such that the Total Variation distance between $A$ and $A'$ obeys 
\begin{align}
    d_{\rm TV}(A, A') \propto |\phi(\hat Y) - \phi(\hat Y')|
\end{align}    
\end{assumption}
Under this assumption, we can prove the following 
\begin{lemma}
    \label{lemma calibration selection}
    Let $S_n = \phi(\hat Y_n)$, $n=1, \dots, N+1$ be defined as in Assumption \ref{assumption ordinal} and assume $BC(Y_n, \hat Y_n) \sim {\cal N}(S_n, 1)$ for all $n=1, \dots, N+1$ with $S_{n} = s_1$ for $n=1, \dots, \bar n$, $S_{n} = s_2 > s_1$ for $n=\bar n+1, \dots, N$, and $S_{N+1} = s_3 > s_2$
    Then, if $|s_3 - s_1|< \frac{1}{40}$, and 
    \begin{align}
        |s_3 - s_1| > \frac{N-n}{N-\bar n+1} |s_2 - s_1|
    \end{align}
    the validity gap defined in \eqref{approximate validity} is reduced by discarding the first $\bar n$ samples.
\end{lemma}
\begin{small}
{\bf Proof of Lemma \ref{lemma calibration selection}}
Let $A_n = BC(Y_n, \hat Y_n)$, $n=1, \dots N+1$.
Under the Lemma assumptions, the validity gap of \cite{barber2023conformal} becomes
\begin{align}
    {\rm gap} = \frac{\sum_{n=1}^N w_n d_{TV}(A_n, A_{N+1})}{1 + \sum_{n=1}^N w_n} =  \frac{\gamma}{2} 
 \frac{\bar n|s_1-s_3| + (N-\bar n)|s_2-s_3|}{N+1}
\end{align}
where $\gamma = \frac15$ as shown in \cite{devroye2018total}.
The claim is obtained by requiring 
\begin{align}
    \frac{\bar n}{N+1}|s_1-s_3| + \frac{\bar n}{N+1}|s_2-s_3| < \frac{N - \bar n}{N - \bar n+1}|s_1-s_2| 
\end{align}
$\square$
\end{small}    

In the experiments, we let $S_n = s(Z_n)$ be the number of qubits in the $n$-th circuit, $n=1, \dots, N+1$ and assume the total variation distance between the corresponding BCs grows linearly in $|S_n - S_{n'}|$.
In this case, the validity gap can be reduced if 
\begin{align}
\label{empirical ordering}
    \frac{\sum_{n=1}^N |S_{n} - S_{N+1}|}{1 + N} >  \min_{s_{min} \in {\mathbb R}} \frac{\sum_{n=1}^N {\bf 1}(S_n > s_{min}) |S_{n} - S_{N+1}|}{1 + \sum_{n=1}^N {\bf 1}(S_n > s_{min})}
\end{align}
While searching for $s_{min} \in {\mathbb R}$ requires trying $N+1$ values and is always feasible, $s_{min}$ is not guaranteed to fulfil \eqref{empirical ordering} if the range of available circuit sizes is limited, however. In our experiments, we let $s_{min}$ be the second largest circuit size in the calibration set.
Calibration samples associated with smaller sizes are discarded or used to fit the shift model described in the next paragraph.

\subsubsection{Calibration training}
\label{section calibration training}
More explicit guarantees can be obtained under 
\begin{assumption}
\label{assumption function}
    Let $\hat Y_{m} \sim P_{\hat Y}$ and $Y_m \sim P_{\hat Y}$, $m=1\dots, M$, be the noisy and noiseless i.i.d. outputs of a QC device and $\hat{BC}(Y, \hat Y)$ the corresponding empirical estimation of \eqref{bhattacharyya coefficient}.
    Then, for any $\hat Y$ and $Y$, there exists a scalar function, $bc = bc(\hat Y)$, that approximates $\hat {BC}(Y, \hat Y)$ and obeys \
    \begin{align}
        \varepsilon_{BC} = | bc(\hat Y) - \hat{BC}(Y, \hat Y) | \ind (\hat Y, Y)
    \end{align} 
    i.e. has attribute-independent residuals. 
\end{assumption}
In this case, we can prove the following 
\begin{theorem}
\label{theorem function}
Let $\hat{BC}$ and $bc$ be as in Assumption \ref{assumption function}, $\hat Y_{nm} \sim P_{\hat Y_n}$ and $Y_{nm} \sim P_{\hat Y_n}$, $n=1, \dots, N+1$, $m=1\dots, M$, be the noisy and noiseless outputs of $N+1$ QC devices, and $\hat {BC}_n$, $n=, \dots, N$, the associated empirical estimation of \eqref{bhattacharyya coefficient}.
Then, for any $t>0$ and any $\alpha \in (0, 1)$, 
\begin{align}
    &{\rm Prob}\left(BC(Y_{N+1}, \hat Y_{N+1}) \geq bc(\hat Y_{N+1}) - q_\alpha - t\right) \geq (1 - \alpha) (1 - 2 e^{-\frac{2 t^2}{M \ C_{N+1}}} ) \\
    & q_\alpha = \inf_q \left\{ \sum_{n=1}^N {\bf 1}\left(|\hat {BC}_n - bc(Y_n, \hat Y_n)| < q\right) \geq \lceil(1 - \alpha)(N+1)\rceil  \right\} 
\end{align}
where $C_{N+1}=|c_{N+1} - c_{N+1}^{-1}|$, $c_{N+1} = \sqrt{\min\{ \min_y P_{Y_n}(y), \min_y P_{\hat Y_n}(y)\} }$.
\end{theorem}

{\bf Remark.} Assuming there exists a conditional model of a circuit's BC with $\hat Y$-independent residuals is a weaker assumption than assuming the existence of the QEM function $\epsilon_{noise}$ in \eqref{noisy outputs}.   

\begin{small}
{\bf Proof of Theorem \ref{theorem function}.}
By assumption, $|\hat{BC}_n - bc(\hat Y_n)|$ are i.i.d. and then exchangeable. 
The definition of $q_\alpha $ implies 
\begin{align}
\label{proof cp bound}
    &{\rm Prob}\left(\hat {BC}_{N+1} \geq bc(\hat Y_{N+1}) - q_\alpha \right) \geq 1 - \alpha
\end{align}
for all $\alpha \in (0, 1)$.
For any $n=1, \dots, N+1$, $BC(Y_n, \hat Y_n)$ is the expectation under $P_{\hat Y_n}$ of $R_n(y) = \sqrt{\frac{P_{Y_n}(y)}{P_{\hat Y_n}(y)}}$.
Hoeffding's inequality bounds the deviation of its empirical estimation in terms of the number of samples, $M$, and the variable range with high probability.
In particular, we have 
\begin{align}
&R_n \in  \left[\sqrt{\min_y P_{Y_n}(y)},  \frac{1}{\sqrt{\min_y P_{\hat Y_n}(y)}}\right] \subseteq [\sqrt{c_n},  \sqrt{1/c_n}]  \\
&c_n = \min\{\min_y P_{Y_n}(y), \min_y P_{\hat Y_n}(y) \}
\end{align}
and, for any $t > 0$, 
\begin{align}
\label{proof concentration bound}
&{\rm Prob} \left( \left|M^{-1} \sum_{m=1}^M \sqrt{\frac{P_{Y_n}(\hat Y_{nm})}{P_{\hat Y_n}(\hat Y_{nm})}} - {\rm E}_{\hat Y_n} R_n(\hat Y_n) \right| \leq t \right) = \\
& = {\rm Prob} \left(\left| \hat BC_n - BC(Y_n, \hat Y_n) \right| \leq t \right) 
\geq  1 - e^{-\frac{2 t^2}{M \ C_n}}
\end{align}
where $C_{n} = |c_{n} - c_{n}^{-1}|$. 
The statement is then obtained by combining the concentration and CP bounds, \eqref{proof concentration bound} and \eqref{proof cp bound}.
$\square$      
\end{small}
The BC approximator needs to be trained on data that are not used to calibrate the CP algorithm.
Ideally, the training data set should be a random subset of all available data, but this is not required for the validity of \eqref{approximate validity}.
In general, however, the empirical estimation of $bc(\hat Y)$ will not fulfil Assumption \ref{assumption function}.
If we assume the BC is Gaussian distributed around a conditional mean that depends only on (features extracted from) $\hat Y$, and $bc$ is a good enough approximation of such conditional mean, 
we do not need Assumption \ref{assumption function} to show that using $\tilde A_n=|BC(Y_n, \hat Y_n) - bc(X_n)|$ instead of $A_n=BC(Y_n, \hat Y_n)$ reduces the validity gap.   

\begin{lemma}
    \label{lemma shift}
    Let $BC_n = BC(Y_n, \hat Y_n) \sim {\cal N}(\mu_n, 1)$, $n=1, \dots, N+1$, where the unknown conditional mean depends only on $\hat Y_n$, i.e. $\mu_n = \mu(\hat Y_n)$. 
    Let $bc(\hat Y_n) \approx {\rm E}(BC_n|\hat Y_n) = \mu_n$ be a pre-trained approximation of the conditional expectation of $BC_n$. 
    Then replacing $A_n=BC_n$ with $\tilde A_n=|BC_n-bc(\hat Y_n)|$ in \eqref{approximate validity} is reduces the validity gap if 
    \begin{align}
        \max_{n} \{|\mu_n - bc(\hat Y_n)|\} \leq \frac{1}{5} \frac{1}{N} \sum_{n=1}^N |\mu_n-\mu_{N+1}|, 
    \end{align} 
\end{lemma}
{\bf Proof of Lemma \ref{lemma shift}.}
According to Theorem 1.3 of \cite{devroye2018total}, the Total Variation distance between two one-dimensional Guassians with unit variance and means $\mu$ and $\mu'$ obeys 
\begin{align}
    \frac{4}{200} |\mu-\mu'| \leq d_{\rm TV}({\cal N}(\mu, 1), {\cal N}(\mu', 1)) \leq \frac{1}{2} |\mu-\mu'|
\end{align}
Consider a CP algorithm calibrated using $\tilde A_n = BC_n-bc(\hat Y_n) \sim {\cal N}(\mu_n-bc(\hat Y), 1)$, $n=1, \dots, N+1$, instead of $A_n = BC_n$.
In this case, the validity gap defined in \eqref{approximate validity} obeys
\begin{align}
    \tilde {\rm gap} &= \frac{1}{N+1} \sum_{n=1}^N d_{\rm TV}(\tilde A_n, \tilde A_{N+1}) \\ 
    & = \frac{1}{N+1} \sum_{n=1}^N d_{\rm TV}({\cal N}(\mu_n-bc(\hat Y_n), 1), {\cal N}(\mu_n-bc(\hat Y_n), 1)) \\
    & \leq \frac{1}{N+1} \frac12 \sum_{n=1}^N |\mu_n-bc(\hat Y_n)-\mu_{N+1}+bc(\hat Y_{N+1})| \\
    & \leq \frac{N}{N+1}  \max_{n} \{|\mu_n - bc(\hat Y_n)| \}
\end{align}
The validity gap of a CP algorithm calibrated using $A_n=BC_n\sim {\cal N}(\mu_n, 1)$ obeys
\begin{align}
    {\rm gap} &= \frac{1}{N+1} \sum_{n=1}^N d_{\rm TV}(A_n, A_{N+1}) \\ 
    & = \frac{1}{N+1} \sum_{n=1}^N d_{\rm TV}({\cal N}(\mu_n, 1), {\cal N}(\mu_n, 1)) \\
    & \leq \frac{1}{N+1} \frac15 \sum_{n=1}^N |\mu_n-\mu_{N+1}|
\end{align}
The latter is larger than $\tilde{\rm gap}$ if 
\begin{align}
    \frac{1}{N} \frac15 \sum_{n=1}^N |\mu_n-\mu_{N+1}| \geq \max_{n} \{|\mu_n - bc(\hat Y_n)| \}
\end{align}
The claim follows from noting that the total variation between two random variables, $Z, Z'$, is larger than the total variation distance between any function applied to $Z$ and $Z'$, e.g. between $|Z|$ and $|Z'|$. 
See the discussion below Lemma 1 of \cite{barber2023conformal} for a discussion about the tightness of the bound.
$\square$

After training $bc$ and calibrating a CP algorithm with $|A_n-bc(\hat Y_n)|$, the prediction intervals for the BC become
\begin{align}
\label{shifted prediction intervals}
        &C_{N+1} = [bc(\hat Y_{N+1}) - q_\alpha, bc(\hat Y_{N+1}) + q_\alpha] \\
        &q_\alpha = {\rm inf}_q \left\{ \sum_{n=1}^N {\bf 1}(|BC_{n} - bc(\hat Y_n)| \leq q) \geq \lceil(1 - \alpha)(N + 1)  \rceil \right\} 
\end{align}
and, under the assumptions of Lemma \ref{lemma shift}, have approximate $(1-\alpha)$ coverage 
\begin{align}
    {\rm Prob}(BC(Y_{N+1}, \hat Y_{N+1}) \in C_{N+1}) \geq (1-\alpha) - \max_{n} \{|\mu_n - bc(\hat Y_n)|\}
\end{align}
In the experiments, we use $bc(Z) = {\rm arg} \min_g {\rm E} \| A - g(Z) \|^2$, where $Z = (\hat{\rm E}(\hat Y), \hat{\rm E}(\hat Y \otimes \hat Y^T))$ are the first and empirical second moments of $\hat Y_n$.

\section{Experiments}
We ran a synthetic experiment to assess the accuracy of the proposed BC estimator and used simulated and real quantum hardware data to validate the prediction of the CP algorithm.
A summary of how we applied the methods described in Section \ref{section extrapolation} in practice is given in the following section \ref{section pipeline}.
\subsection{Pipeline}
\label{section pipeline}
Given a quantum machine, e.g. {\tt ibm\_kyiv}  at \cite{ibm_machines}, we proceed as follows.
\begin{enumerate}
    \item Use the quantum machine to run $N$ classically tractable circuits and collect the corresponding noisy outputs, $\hat Y_{nm}$, $n=1, \dots, N$ and $m=1, \dots M_{shots}$.
    \item Use a classical simulator to run the same $N$ circuits and collect the corresponding noise-free outputs, $Y_{nm}$, $n=1, \dots, N$ and $m=1, \dots M_{shots}$.
    \item Compute the BC for all $n=1, \dots, N$.
    \item Split the data set according to the circuit sizes, $s_n$, e.g. we let $I_{train}=\{n: s_n < s_{max} \}_{n=1}^N$, $s_{max} = {\rm arg} \max_{n} \{s_n\}_{n=1}^N$ and $I_{cal} = \{n: s_n = s_{max} \}_{n=1}^N$.
    \item Use $\{ \hat Y_{n}, Y_{n}\}_{n \in I_{train}}$ to train the shift function, $g(\phi(\hat Y))$, by minimizing $\sum_{n\in I_{train}}(A_n - g(\phi(\hat Y_n)))^2$, where $\phi: P_{\hat Y} \to {\mathbb R}^d$ is an arbitrary feature map. 
    \item Evaluate the transformed conformity scores, $B_{n} = {\rm BC}(Y_n, \hat Y_n) - g(\phi(\hat Y_n))$ for all $n \in I_{cal}$.
    \item Let $Q_\alpha$ be the $1-\alpha$ empirical quantile of the calibration scores, $\{ B_{n}\}_{n\in I_{cal}}$.
    \item Let $\hat Y_{N+1}$ be the noisy output of a classically intractable test circuit and evaluate $g(\phi(\hat Y_{N+1}))$. 
    \item Return the upper bound for the BC between $\hat Y_{N+1}$ and its ideal (unavailable) counterpart,
    \begin{align}
        {\rm BC}(Y_{N+1}, \hat Y_{N+1}) \geq g(\phi(\hat Y_{N+1}) + Q_\alpha) 
    \end{align}
    which hold with probability $1 - \alpha - {\rm gap}$, with 
    \begin{align}
    &{\rm gap} = 2 \frac{\sum_{n\in I_{cal}} d_{\rm TV}(A_n - g(\phi(\hat Y_n)), A_{N+1} - g(\phi(\hat Y_{N+1}))}{|I_{cal}| + 1}  \\    
    &A_n={\rm BC}(Y_n, \hat Y_n) \\
    & \phi(\hat Y_n) = (\hat{\rm E}(\hat Y_n), \ \hat{\rm E}(\hat Y_n \otimes \hat Y_n^T)) = 
    \left( \sum_{m=1}^M\hat Y_{nm}, \ \sum_{m=1}^M \hat Y_{n,} \otimes \hat Y_{nm}^T\right)
    \end{align}
\end{enumerate}

\subsection{Density-ratio estimator}
\label{section logistic regression}
In all experiments, we use a Logistic Regression (LR) parametric estimator for $\hat r = \frac{P_{Y}}{P_{\hat Y}}$, 
\begin{align}
    &\hat r_{\rm LR}(\hat Y_n) = \frac{f(\hat Y_n)}{1 - f(\hat Y_n)}, \\
    &f(Y) = \sigma(\hat \theta^T\phi(Y)) \sim {\rm Prob}(Y \sim P_{Y}|Y)
\end{align}
where $\theta \in \mathbb R^d$, $d \in {\mathbb N}_+$ is an optimized free parameter and $\phi(\hat Y_n)$ a fixed feature map from $\{\hat Y_{n1}, \dots, \hat Y_{nM_{shots}}\}$ to ${\mathbb R}^d$.
Other options include matching all moments of $P_{Y}$ and $r P_{\hat Y}$ under the assumption that the distributions belong to a Reproducing Kernel Hilbert Space (RKHS) \citep{sugiyama2012density}.

\subsection{Synthetic data}
We generated a set of multivariate Bernoulli product distributions of dimension, with dimensions $s \in \{10, 20, 40, 80\}$ and ground truth and perturbed weights, $w, \hat w \in [0, 1]^s$.
Ground truth weights, with profile $w_{log \ i}\propto \frac{\log(1 + i)}{s}$, $w_{rand \ i}\propto \frac{U * i}{s}$, $U \sim {\rm Uniform}_[0, 1]$, and $w_{cos \ i}\propto \cos^2(\frac{\pi}{\epsilon + i})$, $\epsilon=10^{-4}$, $i=1, \dots, s$, were perturbed using three different perturbations, $\epsilon_{log \ i} \propto \log(\epsilon + w_i)$, $\epsilon_{rand \ i} = w_i * V$, $V\in {\cal N}(0, 1)$, and $\epsilon_{cos\  i} = \cos(w_i * 2\pi)$. 
The unspecified proportionality constants guaranteed that all $w_i$ and $\hat w_i$ were nonnegative and lower than 1.
Each ground truth profile was perturbed with all three perturbations.
For all $(w, \hat w)$ pairs, we estimated the Bhattacharyya distance, $d_{\rm BC} = -\log {\rm BC}$, the Kullback–Leibler divergence, $d_{\rm KL}$, and the Total Variation distance, $d_{\rm TV}$ defined in \eqref{total variation}, between the associated ground truth and perturbed distributions with the density-ratio estimator of Sections \ref{section estimating} and \ref{section logistic regression} and compared them with the theoretical Bhattacharyya distance between two multivariate Bernoulli product distributions,
\begin{align}
\label{Bhattacharyya distance}
\bar d_{\rm BC} &= -\log \prod_{i=1}^s \left(\sqrt{w_{i} \hat w_i} + \sqrt{(1- w_{i})(1-\hat w_i)} \right)
\end{align}
In the first run, the highest and lowest BC distances were 1.112 and 0.016, for $w=w_{cos}$, with $\epsilon=\epsilon_{log}$ and $s=80$, and $w=w_{log}$, with $\epsilon=\epsilon_{cos}$ and $s=10$.
\begin{figure}
    \centering
    \includegraphics[width=0.9\linewidth]{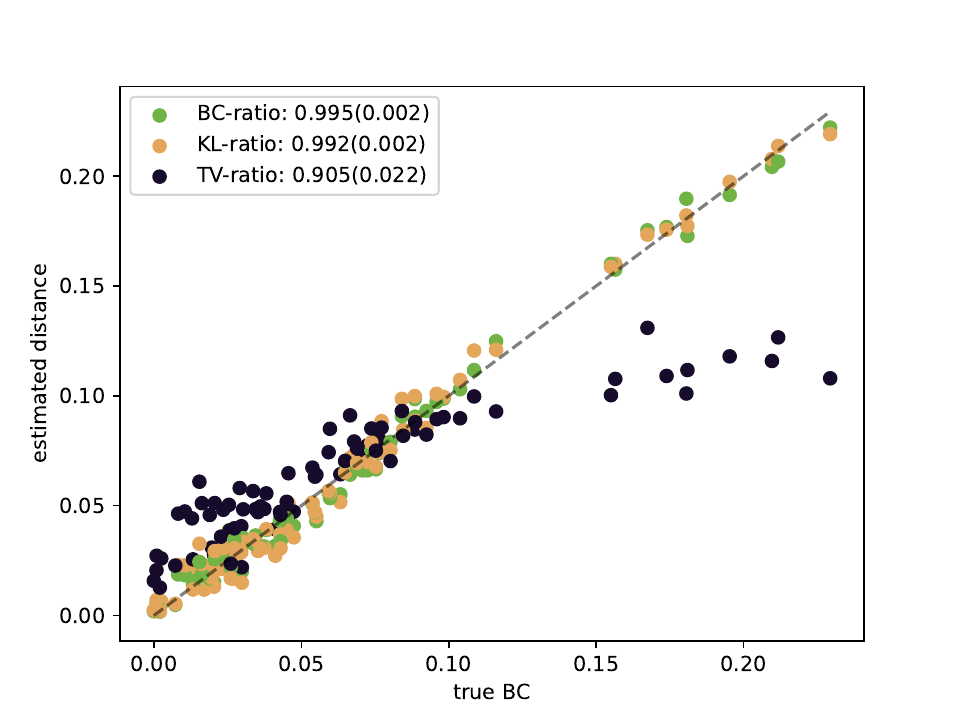}
    \caption{
    Scatter plot between the theoretical Bhattacharyya distance, $\bar d_{\rm BC}$ defined in \eqref{Bhattacharyya distance}, and the density-ratio estimation of $d_{\rm BC}$, $d_{\rm KL}$, and $d_{\rm TV}$ obtained for all weights in the first of 5 simulations.
    All distances were min-max normalized before computing the correlation.
    The legend reports the average correlation and corresponding standard deviations on the 5 equivalent runs.}
    \label{figure bernoulli}
\end{figure}

\subsection{Simulated data}
\label{section IBM simulator}
A \emph{realistic} data set was generated by designing circuits of varying sizes and depths and running them on the IBM \emph{fake backends} and the Qiskit Aer simulator \citep{aer} to obtain the corresponding noisy and ideal outputs, $\hat Y_n$ and $Y_n$. 
36 circuits, called 
{\tt W-State},
{\tt Portfolio Optimization with VQE},
{\tt Deutsch-Jozsa}, 
{\tt Graph State}, 
{\tt GHZ State}, and 
{\tt Variational Quantum Eigensolver}
were generated using the algorithms of \cite{apak2024ketgptdatasetaugmentation} with 6 different sizes $s \in \{5, 7, 9, 11, 13, 15\}$.
12 circuits, called {\tt Random} and {\tt Deep Random}, 
were generated using the Qiskit random generator with parameters, $({\tt nqubits}=s, {\tt depth}=s)$ and $({\tt nqubits}=s, {\tt depth}=3 * s)$, $s \in \{5, 7, 9, 11, 13, 15\}$.
6 additional circuits, all with size 4 but different depths, were generated using the {\tt walker} method of \cite{martina2022learning} with depth parameter $s \in \{5, 7, 9, 11, 13, 15\}$. 
Figure \ref{figure walker} displays an example of a {\tt walker} circuit with depth 3.
All circuits were run 5 times on 5 IBM fake backends, {\tt fake\_cusco}, {\tt fake\_kawasaki}, {\tt fake\_kyiv}',  {\tt fake\_kyoto}, and {\tt fake\_osaka}, and using Qiskit Aer simulator.
All simulations produced $M_{shots}=1000$ quantum shots per run.
We designed four different setups to test the mitigation strategies of Section \ref{section extrapolation}, {\tt all}, {\tt mondrian}, {\tt shift}, and {\tt shift+mondrian}. 
In {\tt shift+mondrian}, we followed the procedure described in Appendix \ref{section pipeline}, where the shift model, $g$, was a Scikit-Learn Random Forest regressor with default parameters, $\phi(\hat Y) = [{\rm E} \hat Y, {\rm vec}({\rm E}(\hat Y - {\rm E} \hat Y) \otimes (\hat Y - {\rm E} \hat Y))]$, and $A \in \{d_{\rm BC}, d_{\rm KL}, d_{\rm TV}\}$. 
In {\tt all} and {\tt mondrian}, we did not train any shift model and used data from circuits with sizes $s<s_{max} = 15$ ({\tt all}) and $s=s_{2nd\ max} = 13$ ({\tt mondrian}) for calibration.
In {\tt shift} and {\tt shift+mondrian}, we trained the shift model on a randomly selected half of all $s<s_{max}$ data ({\tt shift}) and only on $s<s_{2nd \ max}$ data ({\tt shift+mondrian}).
In both cases, we let the calibration set contain all the remaining $s<s_{max}$ data.
The test set always consisted of all $s=s_{max}=15$ data.
Table \ref{table IBM} shows the average size and coverage of the CP prediction intervals for all setups.
\begin{table}[]
    \centering
    \begin{tabular}{|l|c|c|}
    \hline
    $d_{\rm KL}$ (simulated data) & coverage & size\\
    \hline
    {\tt all} & 0.9 (0.0136) & 1.971 (0.604)\\
    {\tt mondrian} & 0.922 (0.0111) &3.273 (1.222)\\
    {\tt shift} & 0.883 (0.0368) &1.492 (0.756)\\
    {\tt shift+mondrian} & 0.905 (0.0222) & 2.816 (0.971)\\
    \hline
    \end{tabular}
    \vspace{.5cm}
    
    \begin{tabular}{|l|c|c|}
    \hline
    $d_{\rm BC}$ (simulated data) & coverage & size\\
    \hline    
    {\tt all} & 0.9 (0.0136) &0.902 (0.297)\\
    {\tt mondrian} & 0.922 (0.0111) & 1.622 (0.604)\\
    {\tt shift} & 0.888 (0.017) & 0.891 (0.445)\\
    {\tt shift+mondrian} & 0.905 (0.022) & 1.413 (0.568)\\
    \hline
    \end{tabular}

    \vspace{.5cm}
    \begin{tabular}{|l|c|c|}
    \hline    
    $d_{\rm TV}$ (simulated data)& coverage & size\\
    \hline    
    {\tt all} & 0.911 (0.032) & 0.934 (0.070)\\
    {\tt mondrian} & 0.988 (0.0136) & 1.004 (0.035)\\
    {\tt shift} & 0.922 (0.040) & 0.851 (0.127)\\
    {\tt shift+mondrian} & 0.961 (0.013) & 0.870 (0.071)\\
    \hline
    \end{tabular}

    \caption{Coverage and size of the CP upper bound on simulated data. The KL, BC, and TV divergences were estimated using the empirical approximation of Section \ref{section estimating} with the Logistic Regression density-ratio estimator described in Section \ref{section logistic regression}. 
    The reported values are the means and standard deviations of 5 leave-one-out runs obtained by removing one machine from all data sets.}
    \label{table IBM}
\end{table}

\subsection{Real data}
\label{section real data} 
    Finally, we tested the algorithms on hardware data from \cite{martina2022learning}. 
    We selected data generated by running 9 modular 4-qubit {\tt walker} circuits of increasing depth on different quantum machines.
    Figure \ref {figure walker} shows a {\tt walker} circuit of depth 3.
    The others were obtained by repeating its structure 2 and 3 times (see \cite{martina2022learning} for more details).
    Each circuit ran on 5 different quantum machines, {\tt ibm\_athens}, {\tt ibm\_casablanca}, {\tt ibm\_lima}, {\tt ibm\_quito}, and {\tt santiago}, with $M_{shots}=1000$ quantum shots per execution.
    The corresponding noise-free data were generated using Qiskit's Aer simulator \citep{aer}.
    We followed the procedure described in Appendix \ref{section pipeline} and considered the same distribution distances and setups of Section \ref{section IBM simulator} with $s$ replaced by the circuit depths.
    The algorithms were all tested on data generated by the deepest circuit (${\rm depth}=9$).
    Table \ref{table martina} shows the average size and coverage of the CP prediction intervals for all setups.
\begin{table}[]
    \centering
    \begin{tabular}{|l|c|c|}
    \hline
    $d_{\rm KL}$ (real data) & coverage & size\\
    \hline
    {\tt all} & 0.787 (0.063) & 0.104 (0.007) \\
    {\tt mondrian} & 0.987 (0.024) & 0.156 (0.004)\\
    {\tt shift} & 0.725 (0.151) & 0.102 (0.017)\\
    {\tt shift+mondrian} & 0.962 (0.049) & 0.161 (0.011)\\
    \hline
    \end{tabular}
    \vspace{.5cm}
    
    \begin{tabular}{|l|c|c|}
    \hline
    $d_{\rm BC}$ (real data) & coverage & size\\
    \hline    
    {\tt all} & 0.837 (0.084) & 0.031 (0.001)\\
    {\tt mondrian} & 0.887 (0.061) & 0.0356 (0.001)\\
    {\tt shift} & 0.750 (0.142) & 0.0275 (0.003)\\
    {\tt shift+mondrian} & 0.925 (0.072) & 0.040 (0.004)\\
    \hline
    \end{tabular}
    
    \vspace{.5cm}
    \begin{tabular}{|l|c|c|}
    \hline    
    $d_{\rm TV}$ (real data) & coverage & size\\
    \hline    
    {\tt all} & 0.775 (0.093) & 0.395 (0.0107)\\
    {\tt mondrian} & 1.000 (0.000) & 0.562 (0.0154)\\
    {\tt shift} & 0.600 (0.108) & 0.372 (0.041)\\
    {\tt shift+mondrian} & 0.912 (0.050) & 0.503 (0.009)\\
    \hline
    \end{tabular}
    \vspace{.5cm}

    \caption{Coverage and size of the CP upper bounds for the real-data experiment described in Section \ref{section real data}. 
    The reported values are the mean and standard deviation of 5 leave-one-out runs, each one obtained by removing the data from one of the 5 quantum machines}
    \label{table martina}
\end{table}
\subsection{Discussion}
    
    \paragraph{Synthetic data}
    The proposed density-ratio estimation seems to be a good proxy for the theoretical Bhattacharyya distance.
    The differences between $d_{\rm KL}$ and $d_{\rm BC}$ and $d_{\rm TV}$ for higher levels of noise show that $d_{\rm TV}$ ignores features accounted by the other distances.
    The strong similarity between $d_{\rm KL}$ and $d_{\rm BC}$ may come from the choice of factorizable Bernoulli distributions.
    Remarkably, the ratio estimation remains good when the dimensionality of the support increases exponentially (from $s=10$ to $s=80$). 
    
    \paragraph{Simulated data}
    The nearly perfect coverage of {\tt all} is probably due to the 5-fold larger calibration set but suggests that the (simulated) noise distribution is exchangeable across different circuit sizes.
{\tt shift} produced more efficient prediction intervals overall but undercovered in some cases.
{\tt mondrian} was the less efficient on these simulations but improved after combining it with a trained shift model, as in {\tt shift+mondrian}.

    \paragraph{Real data} 
    The consistent undercoverage of {\tt all} shows that noise distributions are more affected by changing a circuit's depth than size.
    The performance in the four different setups followed the same pattern as for simulated data.
    {\tt shift} serious low validity was likely due to the reduced size of the training data set.\footnote{The test Error Rate of $g$ was high on average (data not shown).}
    Except for $d_{\rm TV}$, where {\tt shift+mondrian} was the best model, the performance of {\tt mondrian} and {\tt shift+mondrian} was comparable.  

\subsection*{Impact Statement}
This paper presents work whose goal is to advance the field of Machine Learning and Quantum Computing. 
There are many potential societal consequences of our work, none of which we feel must be specifically highlighted here.
\bibliographystyle{apalike}
\bibliography{myrefs}

\end{document}